\documentclass[letterpaper]{article} 
\usepackage{aaai24}  
\usepackage{times}  
\usepackage{helvet}  
\usepackage{courier}  
\usepackage[hyphens]{url}  
\usepackage{graphicx} 
\usepackage{natbib}  
\usepackage{caption} 
\usepackage{algorithm}
\usepackage{algorithmic}

\usepackage{makecell}
\usepackage{tabularx}
\usepackage{booktabs}
\usepackage{subfigure}
\usepackage{multirow}

\usepackage{threeparttable}

\usepackage{newfloat}
\usepackage{listings}
\usepackage{newfloat}
\usepackage{listings}
\DeclareCaptionStyle{ruled}{labelfont=normalfont,labelsep=colon,strut=off} 
\floatstyle{ruled}
\newfloat{listing}{tb}{lst}{}
\floatname{listing}{Listing}
\title{BDIQA: A New Dataset for Video Question Answering to Explore Cognitive Reasoning through Theory of Mind}\author{
    Yuanyuan Mao\textsuperscript{\rm 1,2},
    Xin Lin\textsuperscript{\rm 1,2},
    Qin Ni\textsuperscript{\rm 3},
    Liang He\textsuperscript{\rm 1,2}
    }

\affiliations{
  
\textsuperscript{\rm 1}
Shanghai Key Laboratory of Multidimensional Information Processing, ECNU, Shanghai, China
\\
  \textsuperscript{\rm 2}Department of Computer Science and Technology, East China Normal
University\\
    \textsuperscript{\rm 3}Key Laboratory of Multilingual Education with AI, Shanghai International Studies University\\
    51215901051@stu.ecnu.edu.cn, xlin@cs.ecnu.edu.cn,niqin@shisu.edu.cn,lhe@cs.ecnu.edu.cn
}

\usepackage{bibentry}

\begin{document}

\maketitle

\begin{abstract}
As a foundational component of cognitive intelligence, theory of mind (ToM) can make AI more closely resemble human thought processes, thereby enhancing their interaction and collaboration with human.
In particular, it can significantly improve a model's comprehension of videos in complex scenes.
However, current video question answer (VideoQA) datasets focus on studying causal reasoning within events, few of them genuinely incorporating human ToM. 
Consequently, there is a lack of development in ToM reasoning tasks within the area of VideoQA.
This paper presents BDIQA, the first benchmark to explore the cognitive reasoning capabilities of VideoQA models in the context of ToM. 
BDIQA is inspired by the cognitive development of children's ToM and addresses the current deficiencies in machine ToM within datasets and tasks.
Specifically, it offers tasks at two difficulty levels, assessing \textbf{B}elief, \textbf{D}esire and \textbf{I}ntention (BDI) reasoning in both simple and complex scenarios. 
We conduct evaluations on
several mainstream methods of VideoQA and diagnose their capabilities with zero-shot, few-shot and supervised learning. 
We find that the performance of pre-trained models on cognitive reasoning tasks remains unsatisfactory. 
To counter this challenge, we undertake thorough analysis and experimentation, ultimately presenting two guidelines to enhance cognitive reasoning derived from ablation analysis.
\end{abstract}

\section{Introduction}
In the attempt to understand the mechanisms of human
for advanced intelligence, cognitive intelligence of AI has gained much attention in recent years, such as affecting computing \cite{poria2017review} and human value alignment \cite{carroll2018overview}. 
Theory of mind (ToM) is the basis of human cognition. It represents a set of cognitive abilities which attribute mental states (beliefs, intentions, desires, knowledge, perspectives, etc.) to others and recognizes that these mental states may differ from one’s own \cite{premack1978does}.
The development of ToM also has fundamental significance for AI cognition development.
It can make AI more closely resemble human thought processes, thereby enhancing their interaction and collaboration \cite{smart2018human, agarwal2021making}. 
In particular, integrating ToM reasoning into video question answering (VideoQA), can significantly improve a model's comprehension of videos in complex scenes \cite{zhong2022video,Zellers_2019_CVPR}.
Since ToM reasoning on VideoQA requires to infer hidden information and relationships related to human understanding with the simultaneous verification of multiple skills as well as integrating visual and auditory information.
\begin{figure}[tb]

\centering
\includegraphics[width = 1.0\linewidth,trim=20 20 8 20,clip]{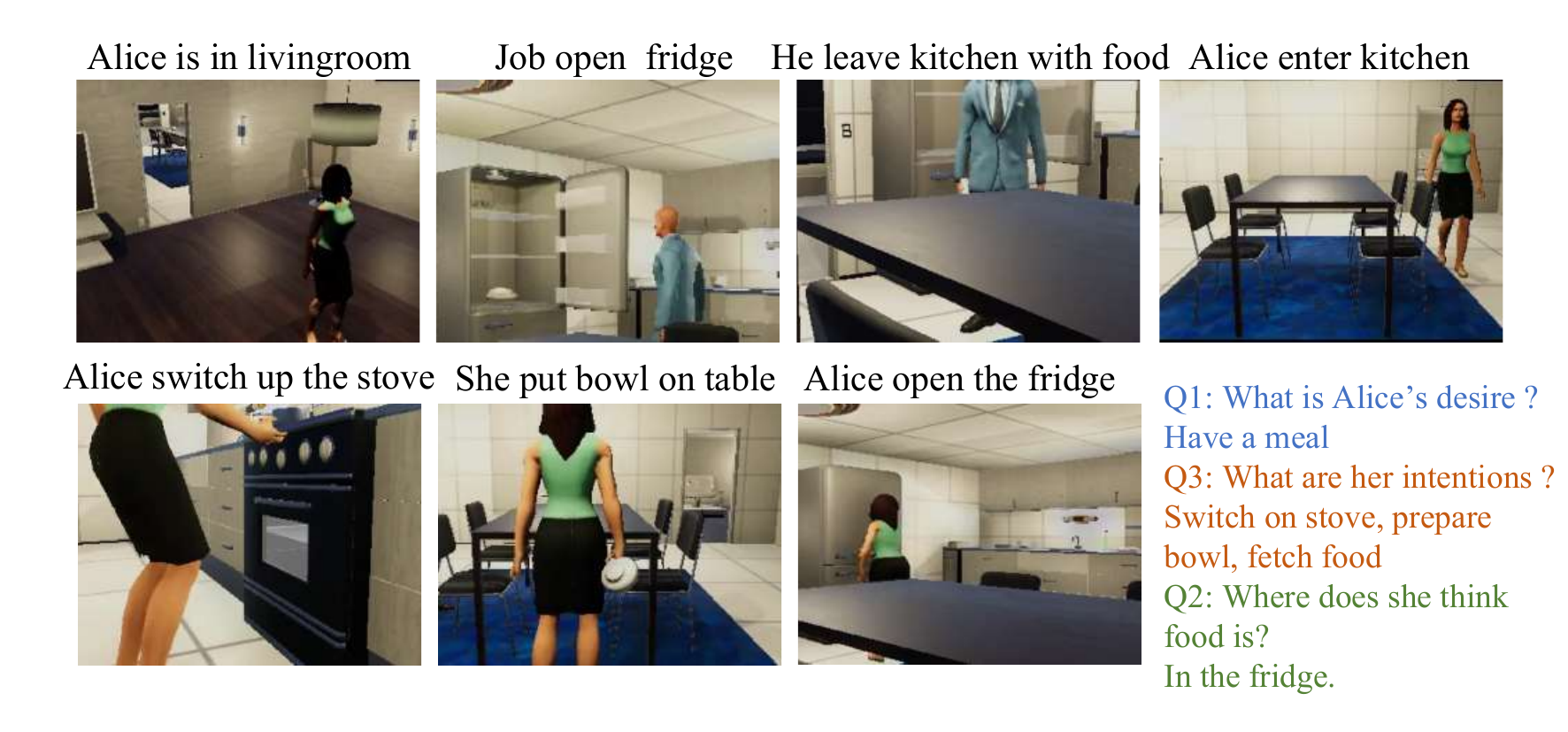}
\caption{An example of how ToM is involved in human action explanation. Job takes the only food in fridge away and leaves the kitchen while Alice is in the living room.
Alice’s desire is to have a meal (desire). In the last picture, she is planning to fetch food (intention). She is walking to the empty fridge because she mistakenly thinks that the food is in the fridge (belief) and hold a false belief about the food. 
}
\label{fig:case2}
\end{figure}

\begin{figure}[t]
    \centering
        \includegraphics[width = 1\linewidth,trim=0 0 70 0,clip]{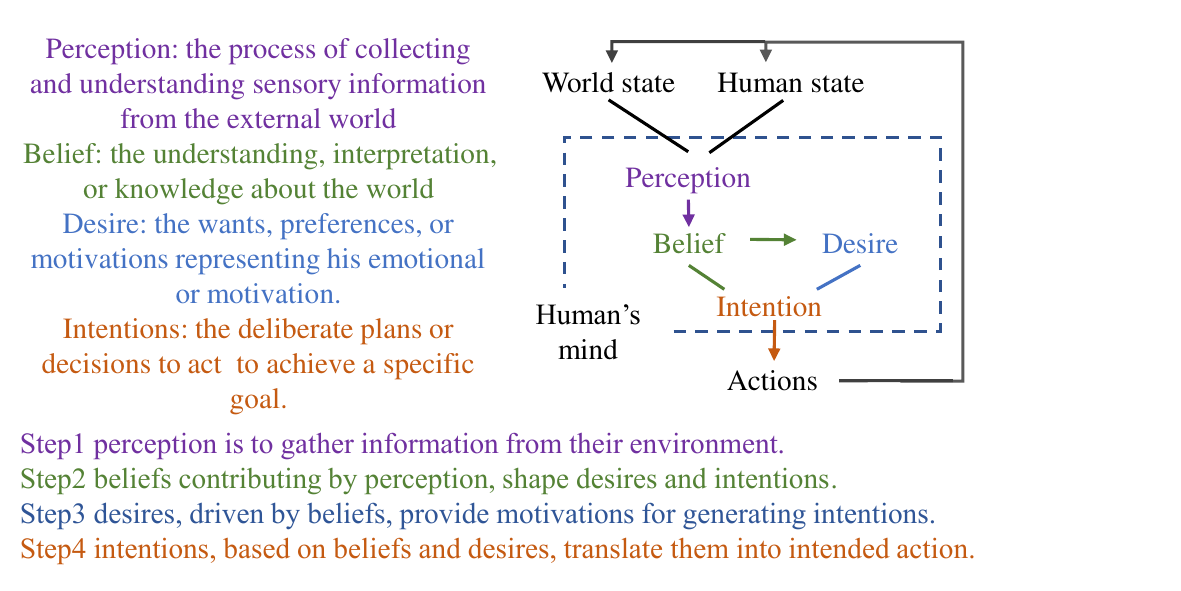}

    \caption{The definition of perception, desire, belief and intention and the relationship of them during human cognitive process.}
    \label{fig:bdi-relationship}
\end{figure}

However, there is a lack of development in cognitive reasoning tasks within the VideoQA domain.
Some of current work on VideoQA study causal reasoning of events but few of them are involved in human mental states.
For example, when asked ``why does Alice walk to the \textbf{empty} fridge?" in Figure \ref{fig:case2}, only by establishing human cognitive processes can models answer this question correctly; most of existing research tend to provide straightforward action descriptions such as wanting to open fridge or fetch food.
 In contrast to descriptions, a comprehensive understanding necessitates an exploration of intrinsic motivation and mechanisms of a complex cognitive process for action generation.
As is shown above, belief, desire, and intention (BDI) of ToM play fundamental roles and can be used to better explain human actions.
Figure \ref{fig:bdi-relationship} and Figure \ref{fig:case2} show the definition of BDI and how humans engage in cognitive reasoning with BDI reasoning.
These three elements work together in a dynamic and complex manner in the human's mind.  
Even though there exists some work \cite{xiao2021next,ko2021air,fang2019intention} for intention understanding in computer vision, we argue that these work don't reveal truly understanding of intentions because they do not explore more behind these actions from a human ToM perspective.


Indeed, there have been prior studies on machine ToM. Nevertheless, persistent challenges remain within this research field, especially concerning tasks and datasets.
In most current work, desire or intention is usually represented as a target or object \cite{baker2017rational_BTOM,gandhi2021baby}. \citeauthor{mao2023review} argue that this setting fails to accurately reflect the real-life experiences of human beings \cite{mao2023review}. And it also lacks joint reasoning on the three concepts of BDI. 
Another limitation is that most datasets are primarily formatted in 2D grids or simple videos \cite {shu2021agent,gandhi2021baby}.
Worthy of reference is that most current work evaluates machine ToM with the children’s ToM \cite{baker2017rational_BTOM, gandhi2021baby}. 
This integration enables AI to mimic and harness the core mechanisms of human development, leading to more sophisticated and adaptable intelligent systems.

Thus, inspired by the development process of children’s ToM, we contribute BDIQA, a benchmark to explore the cognitive reasoning of VideoQA models in \textbf{B}elief, \textbf{D}esire and \textbf{I}ntention of ToM. 
BDIQA fills the gap of cognitive reasoning tasks in VideoQA and also addresses the limitations of current machine ToM, which suffers from lacks of diversity in both dataset format and joint reasoning tasks.
The questions of BDI are involved in human actions, revealing mental states and cognitive processes.
 BDIQA contains a two-level structure at different difficulty for ToM sub-abilities. 
Children's ToM development exhibits unique cognitive abilities, ordered developmental stages, and essential learning processes. BDIQA is leveraged by these characteristics to provide enriching and instructive experiences.

As a pioneering cognitive intelligence task, BDIQA is evaluated on several VideoQA methods such as memory network \cite{gao2018motionComem}, graph neural network (GNN) \cite{duan2022surveyembodies}, modular networks \cite{le2021hierarchicalHCRN} and pre-trained models \cite{yang2021just}
as well as models of video classification \cite{li2021uniformer} with with zero-shot, few-shot and supervised learning. Notably, the current pre-trained models do not exhibit satisfactory performance on BDIQA. Thus, we delve deeper into pre-trained models' exploration on BDIQA and build our cognition reasoning models.
To enable pre-trained models to adapt to cognitive reasoning, we employ leading visual techniques and expand the pre-trained model through the integration of a memory module to proceed multi-step reasoning in complex scences.
Finally, our method outperforms with an overall margin of 1.17\% over the baseline and we derive two cognitive reasoning guidelines through ablation study. 

Our contributions can be summarized as follows: 
\begin{itemize}
    \item We contribute BDIQA, a benchmark to explore the cognitive reasoning of VideoQA in BDI of ToM with a two-level structure to facilitate phased evaluation.
    \item  We extensively analyze the baselines and establish our VideoQA models on BDIQA, providing
detailed results for various question types, and heuristic observations that can guide future research in this area.
\end{itemize}

 \section{Related Work}
\textbf{Machine ToM} 
has been developed with a lot of work to incorporate ToM capabilities into machines \cite{pmlr-v80-rabinowitz18a,ko2021air} but the existing research has some limitations on machine ToM. 
On one hand, the experimental setup and data format are too simple for intention and desire reasoning within a 2D grid where the agents view their desire or intention as a goal with discrete actions \cite{gandhi2021baby, shu2021agent}, which is challenging to apply to real-world situations with humans.
\citeauthor{baker2017rational_BTOM} provided fewer than 100
samples in a 2D grid to facilitate joint inference of desires
and beliefs.
In overcook \cite{overcook}, they collected about 400 trajectories from a two-player cooperative game in 2D grid for intention reasoning and action prediction.
On the other hand,
most datasets of belief reasoning are in isolation from intentions and desires in spite of some of joint inference datasets \cite{baker2017rational_BTOM}.
Some have developed NLP question-answer (QA) datasets formalized ``Sally and Anne" test with a story \cite{grant2017can, nematzadeh2018evaluatingtom-bAbi2, DBLP:conf/emnlp/LeBN19/bAbi3} as well as image-based datasets \cite{eysenbach2016mistaken,duan2022boss}.



\textbf{VideoQA} tasks can be categorized into two types based on questions: factoid VideoQA and inference VideoQA \cite{zhong2022video}. 
Neither of these tasks currently involves human cognitive reasoning within ToM.
Factoid questions directly inquire about visual facts, such as locations or colors \cite{wang2018movie,lei2018tvqa,maharaj2017dataset}. 
On the other hand, inference VideoQA explores the logic within dynamic scenarios \cite{xiao2021next,yi2019clevrer,mao2022clevrer}.
As proposed by recent works, VideoQA currently emphasizes temporal and causal relationships that feature temporal dynamics. 
CLEVRER \cite{yi2019clevrer} and CLEVRER-Human \cite{mao2022clevrer} specially studied temporal and causal relationships of physical motions. NExT-QA \cite{xiao2021next} is the first VideoQA dataset for casual and temporal action reasoning towards deeper explanation. With causal reasoning, VideoQA has been developing in a more intelligent direction while few of these datasets has incorporated human cognitive reasoning within ToM.

Various methods enhance VideoQA via four components: video encoder, question encoder, cross-modal interaction, and answer decoder. 
Video encoders evolve for effective representation, using 2D CNNs (e.g., ResNet \cite{restnet_He}) for appearance and 3D CNNs (e.g., C3D, \cite{c3dtran2015learning}, I3D \cite{carreira2017quoI3D}) for motion. Question encoders use pre-trained models (e.g., GloVe \cite{pennington2014glove}, BERT \cite{devlin2018bert}) for semantic understanding. Sequential models (RNNs \cite{Lev2015RNNFV}, CNNs \cite{Kato2015VisualLM}, Transformers \cite{Vaswani2017AttentionIA}) process vision and language. Cross-modal modules (spatial/temporal attention\cite{hOR_dang2021hierarchical}, co-attention \cite{fan2019heterogeneousHME}, multi-cycle memory \cite{fan2019heterogeneousHME,gao2018motionComem}, GNN \cite{jiang2020HGA,wang2021dualvgrDual}, and conditional relation networks \cite{hOR_dang2021hierarchical,le2021hierarchicalHCRN}) fuse information for reasoning. 
The answer decoder emerges as the linchpin in the VideoQA pipeline to synthesize coherent and contextually appropriate responses. Leveraging the enriched features, the decoder is adept at inferring answers that reflect a profound understanding of the input data.
Recent strides in visual-language pre-training profoundly affect VideoQA. Abundant pre-trained models leverage advanced techniques and transfer learning \cite{yang2021just, ViTradford2021learning, yangzerofrozen}, even adapting from other visual-language tasks \cite{wang2023all,XCLIP}.

\section{Dataset}
With a two-level structure inspired by children's ToM development, we introduce BQIQA, a VideoQA dataset to assess machine ToM of BDI. BDIQA assesses the sub-abilities of machine ToM at each level and for every concept.
A video in BDIQA includes two characters performing household activities.
And our dataset asks questions about characters’ belief, desire and intention as well as perception. We generate videos in the way of animation. Synthetic datasets can control the generation of annotation, and generate large-scale datasets. 
Further, we conduct human evaluation of BDIQA to quantify human reasoning ability on BDIQA.
Finally, by human validation and manual filtering, we obtain 3,527 videos and 19,932 QA pairs.

\subsection{Task Setup}
\begin{figure}[tb]

\centering
\includegraphics[width = 1.0\linewidth,trim=50 30 60 20,clip]{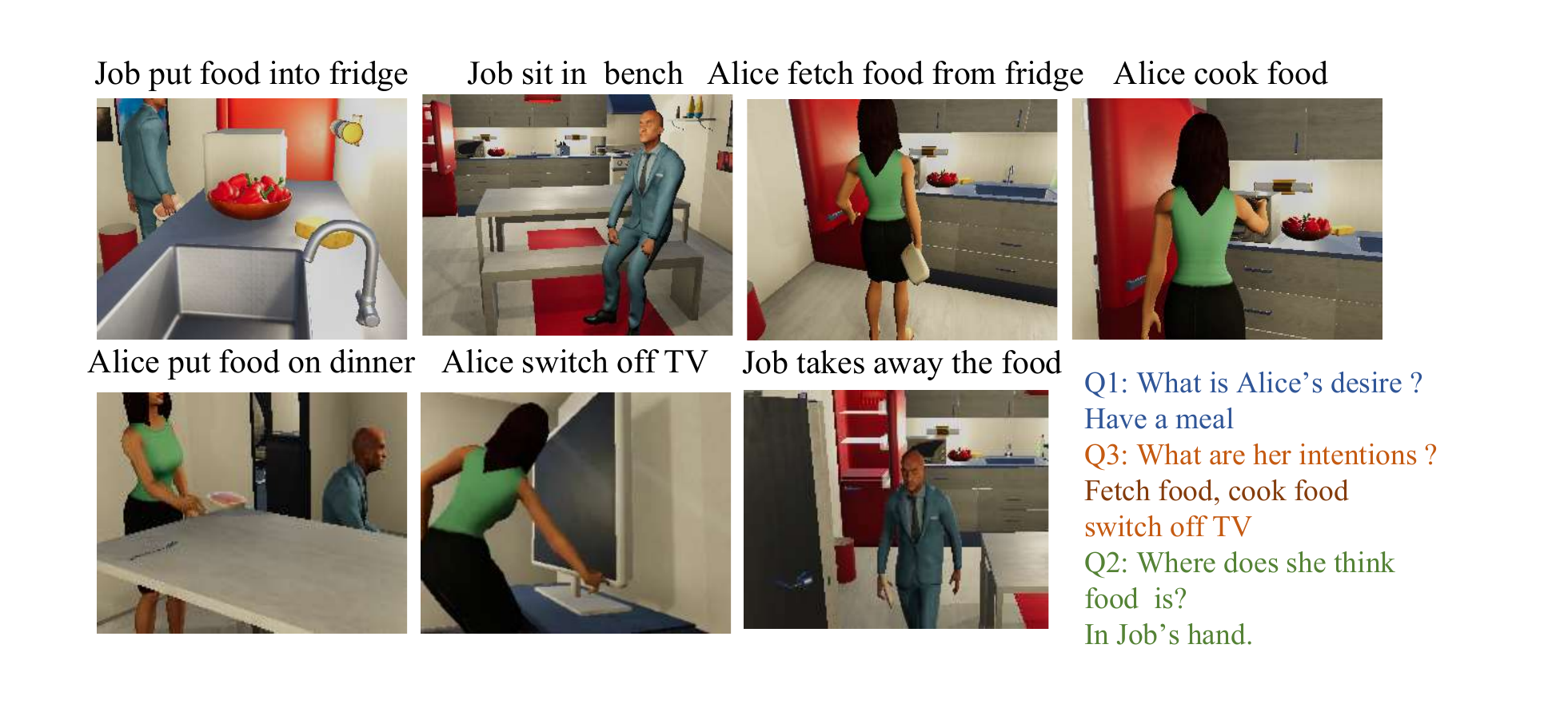}
\caption{A example for true belief and unsatisfied desire for Alice.
Alice fails to have a meal because of Job. And during that time they never leave kitchen and they have a true belief about the food which is consistent with real world.
``Fetch food" is a required sub-task.``Switch off TV" is an optional sub-task because it is a necessary step for ``have a meal".
}
\label{fig:case1}
\end{figure}

There are two characters in the video, a male character called \textit{Job} and a female character called \textit{Alice}. 
For each video, \textit{Alice}'s desire is to complete a household activity and she makes intended plans. 
For the three mental states of BDI, psychology currently subdivides them into multiple ToM sub-abilities from simple to complex tasks.
We build a two-level dataset to examine different aspects of ToM according to these difficulty divisions. There are five types of questions in our BDIQA - belief question, desire question, intention question, ``where" question and ``yes/no" question.
Each mental state question is divided into two levels: \textit{level 0} and \textit{level 1}, where \textit{level 0} is easier than \textit{level 1} in infants.  
At the first level, BDIQA dataset involves simple reasoning tasks with satisfied desires, simple intentions and true beliefs; while at the second level, we set for harder tasks with unsatisfied desires, complex intentions and false beliefs.
And we list sub-abilities of all levels of the three mental states in Table. \ref{tab:sub-abilities} and give examples in our dataset with explanation in detail shown in Figure \ref{fig:case1} and Figure \ref{fig:case2}.

\begin{table}[tb]
\centering\
\renewcommand\arraystretch{0.9}
\setlength\tabcolsep{3pt}   

\scalebox{0.90}{
\begin{tabular}{ccc}
   
\toprule
  
{\textit{level 0:} Desire} & {\textit{level 0:}Intention} &
{\textit{level 0:}Belief}
\\
\midrule

  know that human take  & know the pur-
  & 
  \multirow{1}{*}{know} \\
    action to meet one's desires & suit of goals
  & true beliefs\\
  
  \toprule
  
{\textit{level 1:}Desire} & {\textit{level 1:}Intention} &
{\textit{level 1:}Belief}
\\
\midrule

 know that desires 
 & know the cho- &
  \multirow{1}{*}{know } \\
  are not always satisfied
 & ice of plans
 & {false beliefs} \\
 
  \bottomrule
\end{tabular}%
}
    \caption{Two levels ToM sub-abilities of BDIQA.}
\label{tab:sub-abilities}%

\end{table}

\textbf{Desire} is represented by the household activity which the character in the video wants to complete.
We design 10 major household activities. For each household activity, the character called \textit{Alice} will make plans to complete the household activity. 
In human desire reasoning, ``desire-outcome matching strategy" \cite{schult2002children} refers to a cognitive process in which humans match their desires with the outcomes they expect to achieve. This strategy demonstrates a simple understanding of relationship between actions and satisfied desired outcomes while it can not be expanded to unsatisfied desire. 
For example in Figure \ref{fig:case1} we can only infer Alice's desires from her actions, not the final outcomes.
It is reported that 3-5 years old children show worse performance on unsatisfied desire \cite{schult2002children}.
Therefore, in \textit{level 1} BDIQA, we add a harder situation where Job would make Alice's desire unsatisfied.

\textbf{Intention} is represented by the sub-task which can help to accomplish one's desire.
Although there is still controversy on time, \citeauthor{vaish2005baby} has shown that beginning around 9 months, infants understand others’ actions as driven by goals, and by 12 to 14 months infants understand others’ choice of plans. It indicates that infants understand that humans take perception into account and attend to only a subset of all things before choosing an action plan. 
Therefore, in \textit{level 0}, Alice will carry out only a set of steps for planning; in \textit{level 1}, a more varied set of steps and random sequences actions will be executed. 
To be specific, we divide sub-tasks into two categories: \textit{required sub-task} and \textit{optional sub-task}.
The \textit{required tasks} are directly related to the corresponding household activities. \textit{Optional tasks} are not usually necessary steps but also accord with human commonsense. These optional sub-tasks will increase the difficulty in judging the true desire of the character. 
Because there are one more intentions for one video, intention questions are in the way of that ``what does \textit{char} do after \textit{intention}? ".

\textbf{Belief} is represented by the location of an object which a character thinks. We adapt the ``Sally and Anne" test \cite{wimmer1983beliefs} to our household activity and ask true belief (\textit{level 0}) and false belief questions (\textit{level 1}). Mastering false belief means understanding that beliefs belong to people's minds and may not correspond to the external world which has been taken as the standard measure that children have a ``mentalistic" understanding of beliefs \cite{broekhof2015understanding}. 
The two examples in Figure \ref{fig:case2}, \ref{fig:case1} can help understand true and false beliefs.
In a video, the character may leave the room and result in false belief. 
Following the classic ``Sally and Anny" test , supposing that when a man leaves a room, he mistakenly thinks an object which has moved to another place is still in its original place and causes a false belief for the object. Belief questions can be formatted ``where does \textit{char} think \textit{object} is? ".

\textbf{``Where" questions} are with ``where is the \textit{object}” which is different with the belief questions because the special word ``think". ``Where" question can be classified as perception task.
For each object, our dataset asks the initial location and the last location of objects following \cite{grant2017can}. 
Actually, reasoning for belief questions often requires first to answer ``where" questions in our setting, and then choosing one of the candidate's locations based on the character's trajectory. Therefore, belief reasoning is a multi-step task.

\textbf{Yes/no questions} are added to identify whether the models can judge true belief and false belief. They are templated as ``does \textit{char} have a false/true belief about \textit{object}?". We ask the belief question in the way of ``think" instead of ``belief'', because we believe that the current models do not yet understand the concept of ``belief"  which leads to the wrong answer. ``Yes/no" questions explicitly complement the belief questions.
\subsection{Dataset Construction}
    
We choose \textit{VirtualHome} \cite{puig2020watch_WAH} for us to generate videos. 
Based on objects and actions in \textit{Virtualhome}, we design 10 major household activities and 28 categories of sub-tasks. \textit{VirtualHome} is composed of 50 custom-designed departments and 4 kinds of room (bathroom, living room, kitchen and bedroom) in each department. We first identified specific sub-tasks for each household activity and design 12 templates for each household task for story generation in our videos.

\textbf{Variation of data}
Fifty different scenarios are provided with various layouts, objects with different sizes  and colors that we can create a large set of physical scene. In addition to that, we categorize each object so that the character can interact with a class of objects rather than a single object. For example, when Alice cooks food, she can cook chicken with the stove or heat a cake with the microwave. In addition, Alice may randomly take actions within common sense. 
\\\textbf{Question Generation} 
Since these characters follow scripts to take actions and housework activities that we have designed, the housework activities and sub-tasks naturally become the labels of the desire and intention. For a video, there will be a desire question of Alice. The intention questions are based on consecutive sub-tasks in each video.
In addition, we track the locations of each object and character, and following the rule \citeauthor{grant2017can}, we can get the beliefs of each character about each object at different times. There are two belief questions, two ``yes/no" question about
the two characters, two ``where" questions for the initial location and the last location of each object.
We generate belief questions of objects which one character hold a true belief and the other hold the false belief about.


\subsection{Human Evaluation and Quality Control}
We conduct a crowdsourced evaluation to quantify human cognitive reasoning ability
over BDIQA. We randomly sample round 2,000 QA pairs  with 90\% test set and 10\% train set and design a web interface for data collection online.
Each person is randomly assigned 6 videos and their questions. Each QA pair is assigned to over 3 random annotators.
All human data is filtered based on time spent answering questions and accuracy of certain participant. 
Subsequently, we conduct quality control with expert re-labeling on questions with poor human performance.
More detail can be found in supplementary material\footnote[1]{
Supplementary https://github.com/mao-yy/BDIQA.git
}.
Finally, the human performance reaches at 84.23\% on filtered test set and we compare the results of human and models in Section \ref{sec_exp}. 
\subsection{Data Statistics}
\begin{figure}[tb]
    \centering
    
    \includegraphics[width = 1\linewidth,trim=145 50 97 0,clip]{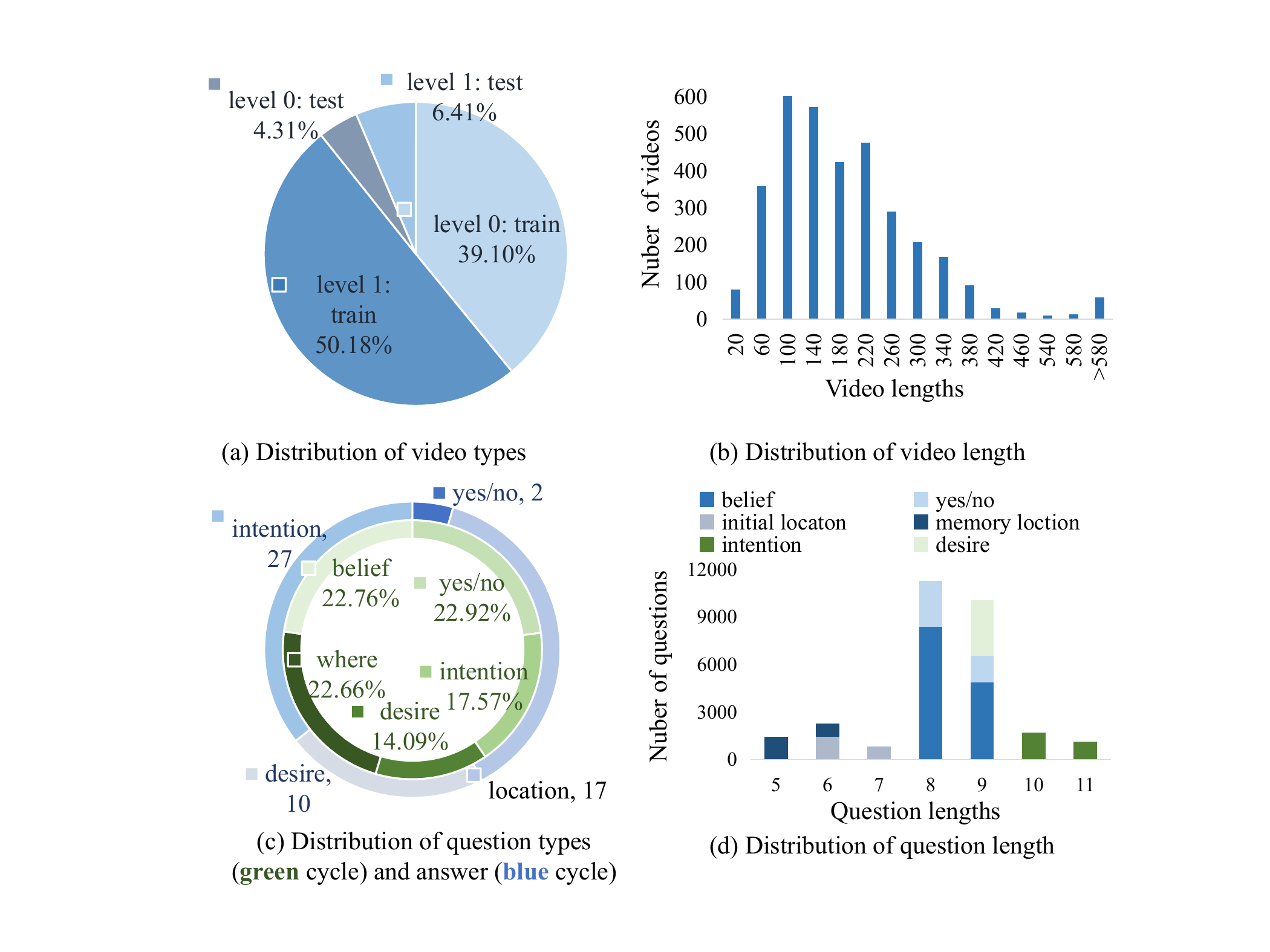}
    

    \caption{Data statistics.}
    \label{fig:data_static-label}
\end{figure}

BDIQA\footnotemark[1]
contains 3,527 videos with 320*240 RGB frames and 19,932 QA pairs, including ~90\% for training, ~10\% for testing. 
Detailed statistics are given in Figure \ref{fig:data_static-label}. 
From Figure \ref{fig:data_static-label}(a) we can see that the number videos of each level accounts for about half of the whole dataset.
And the average video length is about 192 frames and the number of video length in the dataset ranges from 30 to 1000 frames with a large time span which also proposes a challenge of long sequence videos for VideoQA.
As is shown in Figure \ref{fig:data_static-label}(c), BDI questions are the majority, constituting 54\%. ``Yes/no" questions of understanding as auxiliary reasoning for belief questions compose 22\% of the whole dataset. Apart from these, there are 22\% of ``where" questions which focus on describing the locations. There are 58 answers and 6 templates 
to generate questions.
Therefore, in Figure \ref{fig:data_static-label}(d) the distribution of the question length is only from 5 to 11.
On the whole, the questions and answers in ours are simpler than that of the counterparts.
We provide various information about human behaviour, including action scripts, action localization, and scene details. We hope that researchers can use this information to expand our dataset and offer more intricate tasks, thereby exploring a wider range of human mental states. Detailed statistics can be found in supplementary material.


   


\section{Experiments}\label{sec_exp}
This section contains comprehensive experiments and intensive analyses of BDIQA. We conduct evaluations on several mainstream methods of VideoQA models and diagnose their capabilities to deal with different tasks respectively. \
In our primary experiments, we conduct zero-shot, few-shot, and supervised learning experiments on the video models. 
Subsequently, we run all models on both \textit{level 0} and \textit{level 1} datasets to examine the dataset hierarchy.
Unexpectedly, 
Our discoveries unveil a distinct disparity between existing pre-trained models and the end-to-end models when applied to BDIQA.
To address the lack of cognitive reasoning ability exhibited by existing pre-trained models trained solely on existing datasets, we endeavor to improve them with the best visual backbone and memory module. 
And we propose guidances on cognitive reasoning tasks through ablation study to enhance their performance
on BDIQA.
More detailed analysis can be found in supplementary material.

\subsection{Experimental Setup}
\subsubsection{Baselines}
\begin{itemize}
\item \textbf{Memory based models.} 
HME \cite{fan2019heterogeneousHME} introduces dynamic memory network (DMN) for multi-step reasoning.
Comem \cite{gao2018motionComem} follows HME and design co-attention for multi-modal integration.
\item \textbf{Hierarchical Module.}
 HCRN \cite{le2021hierarchicalHCRN} introduces a reusable conditional relation network (CRN) module to obtain the relationship between the various parts of the video at different levels.
\item \textbf{GNN.}
HGA \cite{jiang2020HGA} utilizes a heterogeneous graph reasoning module and co-attention unit to capture local and global correlations between video clips and linguistic concepts, while Dual \cite{wang2021dualvgrDual} employs a stacked, iterative graph-based reasoning unit for multi-step reasoning.
\item \textbf{Pre-trained Model.} Frozen \cite{yangzerofrozen} and JustAsk \cite{yang2021just} are transformer-based models with VideoQA dataset. 
 ClipBERT \cite{lei2021less} is an efficient framework for end-to-end learning for image-text and video-text tasks with sparse sampling.
\end{itemize}

\subsubsection{Setup}
Frames are resized to 224 × 224 as input. We divide a video into 12 clips, and randomly sample 16 frames from each clip.
We employ two visual backbone models.
Follow original version of JustAsk, one is S3D \cite{s3dxie2018rethinking} pre-trained on HowTo100M \cite{miech2019howto100m}.
The other one is a  composed of ResNet-101
 \cite{restnet_He} trained on ImageNet \cite{restnet_He} to extract the per-frame
appearance feature, and 3D ResNeXt-101 \cite{motion_fea_Hara2017CanS3, motion2_xie}
pre-trained
on Kinetics \cite{kay2017kinetics} to extract clip-level motion information. 
On the language side, all models use Bert \cite{devlin2018bert} to tokenize text inputs and get the token embedding.
The initial learning rate is from  1e-4 to 1e-5 with cosine decayed in subsequent epochs. 
Experiments are performed on an NVIDIA 3090 GPU.  
More setup can be found in supplementary material.

Besides overall accuracy (overall \(Acc\)), we also report the question per-type accuracy and the accuracy of different tasks, i.e., cognitive reasoning including all BDI questions (\(Acc_{bdi}\)) and perception reasoning including initial location and memory location question (\(Acc_{p}\)).
\subsection{Main Result}
\subsubsection{Zero-shot on BDIQA}
Table \ref{fig:zero_shot} (left) illustrates the zero-shot performance of video-language models on BDIQA. 
JustAsk which is trained on large-scale automatically generated VideoQA data, outperforms the other methods significantly in both BDI reasoning and perception reasoning. However, we also observe that, Frozen
only achieves approximately 0.40\%  and ClipBERT \cite{lei2021less} only achieves 1.09\% for cognitive reasoning tasks.
\begin{table}[b]
\centering
\scalebox{1}{

\begin{tabular}{c|cc|cc}
\toprule

  \multirow{2}{*}{Models}       & \(Acc_{bdi}\)      &  \(Acc_{p}\)       &  \(Acc_{bdi}\)      &  \(Acc_{p}\) \\
  [-0.15em]
  
  & \multicolumn{2}{c|}{{zero-shot}} & \multicolumn{2}{c}{{few-shot}} \\
        
 \midrule
 ClipBERT   & 1.09\% & 2.05\% & 32.14\% & 40.87\% \\
 JustAsk    & 13.55\%       & 29.98\%       & 16.72\%       & 30.80\%      \\
Frozen  & 0.40\%        & 2.72\%        & 5.04\%        & 10.89\%       \\

\bottomrule
\end{tabular}%
}
\caption{Comparison with baselines for zero-shot and few-shot VideoQA on BDIQA.}

\label{fig:zero_shot}
\end{table}

\subsubsection{Few-shot on BDIQA}
Table \ref{fig:zero_shot} (right) displays the results of few-shot performance on BDIQA with 10\% data in the train set. 
ClipBERT presents great improvement with few-shot VideoQA and shows the best performance among the pre-trained models.
Yet, there is only a slight improvement in Frozen with few-shot VideoQA. JustAsk correctly answers only 16.72\% of BDI questions, in contrast to its higher accuracy on perception reasoning.

Furthermore, we conducted a video classification task on existing video models on desired reasoning. In comparison to kinetics400 \cite{kay2017kinetics} with 400 labels, the desired reasoning task in BDIQA is considerably easier with only 10 labels, and there is some overlap actions between the two datasets.
Table \ref{fig:video_cls} exhibits the results of desired reasoning at two levels. Xclip outperforms other methods, achieving an accuracy of 32.81\%. However, the overall accuracy of VideoMAE and VideoMAEV2 are around 10\%, which are even worse than random asking. Moreover, it is evident that each model performs poorly 
at \textit{level 1} compared to \textit{level 0}.

Overall, relies on training from existing datasets and current pre-trained models, they cannot solve our BDI reasoning task effectively.
In particular, at \textit{level 1} of BDI reasoning, all models can not answer well.

\begin{table}[bt]
\centering
\scalebox{0.9}{
\begin{tabular}{cccc}
\toprule
Models     & desire  & level 0 & level 1 \\
\midrule
Uniformer (\citeauthor{li2021uniformer})& 26.13\% & 28.04\% & 26.65\%  \\
VideoMAE (\citeauthor{tong2022videomae})& 9.14\%  & 9.76\%  & 7.48\%   \\
VideoMAE (\citeauthor{wang2023videomaev2})&10.66\% & 11.85\% & 9.35\%   \\
XClip (\citeauthor{XCLIP})& 32.81\% & 37.38\% & 20.56\%
 \\
\bottomrule
\end{tabular}%

}
\caption{Comparison with pre-trained models of video classification for few-shot on BDIQA desire reasoning.}

\label{fig:video_cls}

\end{table}

\subsubsection{Supervised Learning} 
As is shown in Table \ref{tab:over_res}, Comem \cite{gao2018motionComem} demonstrates the best performance among all methods.
While HGA \cite{jiang2020HGA} exhibits the poorest performance. The inadequate performance of HGA on the desire question task contributes to the overall result.
Dual, a GNN based model and HCRN, a modular networks perform also well in overall accuracy.
The performance of these better models may thank to the structures of multi-hop inference with a fusion of visual and language features.
With the same epoches, we also compare the fine-tuning results of the pre-trained models.
Frozen and JustAsk don't present good results as expected on the two tasks, especially JustAsk, despite its training on numerous VideoQA datasets and prior experience in solving similar tasks. 
It is observed that models with S3D as the video backbone such as Frozen, JustAsk, and HGA, do not perform well on BDIQA.
Thus, the poor performance of the two pre-trained models may be due to the video backbone. 
It is also hypothesized that this outcome can be attributed to the significant disparity between our dataset and the previous VideoQA datasets.

\begin{table}[tb!]
\centering
\scalebox{0.90}{

\begin{tabular}{cccc}
\toprule
Models    & overall \(Acc\)   & \(Acc_{bdi}\)      &  \(Acc_{p}\)   \\
\midrule

\multicolumn{4}{l}{{Visual backbone: ResNet+ResNeXt-101}}\\

Dual (\citeauthor{wang2021dualvgrDual}) & 75.39\%  & 74.78\% & 70.30\% \\
HCRN (\citeauthor{le2021hierarchicalHCRN})   & 76.39\%  & 73.49\% & \textbf{73.02\%} \\
HME (\citeauthor{fan2019heterogeneousHME})   & 71.51\% & 69.05\% & 59.58\% \\
Comem (\citeauthor{gao2018motionComem})   & \textbf{79.84\%}  & \textbf{78.73\%} & 71.93\% \\
\hline
\multicolumn{4}{l}{{Visual backbone: S3D \cite{s3dxie2018rethinking}}}\\
HGA (\citeauthor{jiang2020HGA})    & 58.63\%  & 52.82\% & 63.76\% \\
JustAsk (\citeauthor{yang2021just}) & 64.59\% & 64.29\% & 64.30\%  \\
Frozen (\citeauthor{yangzerofrozen})  & 71.49\%  & 70.03\% & 68.66\% \\
\hline
Human   & -       & 84.23\% & 80.54\% \\
\bottomrule
\end{tabular}%

}
\caption{ A comparison with results of end-to-end models methods and pre-trained models of VideoQA.}
\label{tab:over_res}

\end{table}
\subsubsection{Dataset Validation}
In order to obtain the validity of the dataset hierarchy of BDIQA, we conduct experiments on \textit{level 0} and \textit{level 1} datasets respectively. We provide an intensive comparison of baselines on BDIQA with human performance.
As is shown in Table \ref{level_res} (the best results for \textit{level 0} are \textbf{bolded} and these of \textit{level 1} are \underline{underlined}), human always perform better than models on every question type except the initial location question and belief question of \textit{level 0}. The gap performance of \textit{level 0} and \textit{level 1} on human and models shows that the task of \textit{level 1} is harder than that of \textit{level 0}. 
Despite demonstrating that belief questions are a multi-step task built upon ``where" questions, the major models do not prioritize ``where" questions over belief questions. For example, for Comem \cite{gao2018motionComem} the accuracy of belief question is 81.36\% over that of ``where"  question 
 (73.93\% ) at \textit{level 1}.
One possible explanation for this is the imbalance in answers across different question types. Another factor could be that the models' inference process does not align with our expectations.
Additionally, we discover that the small gap between the performance of the best VideoQA model and human is 5.50\% in cognitive reasoning potentially attributed to the less diversity of questions and answers of BDIQA.


\begin{table*}[bth]
\centering
\scalebox{0.90}{
\renewcommand\arraystretch{1}

\begin{tabular}{ccccccccccc}
\toprule
{Models} &
  {Level} &
  {overall \(Acc\)} &
  {\(Acc_{bdi}\)} &
  {\(Acc_{p}\)} &
  {desire} &
  {intention} &
  {belief} &
  {initial} &
  {memory} &
  {yes/no} \\
  \midrule
\multirow{2}{*}{Dual} &
  l0 &
  82.13\% &
  77.43\% &
  \textbf{76.30\%} &
   77.54\% &
 70.65\% &
  79.76\% &
  88.79\% &
  \textbf{63.46\%} &
   96.22\% \\
 &
  l1 &
  76.76\% &
  74.78\% &
  66.67\% &
  77.21\% &
  75.03\% &
  75.32\% &
  62.34\% &
  \underline{ 70.89\%} &
  \underline{ 91.97\%} \\
  \midrule
\multirow{2}{*}{HGA} &
  l0 &
  69.37\% &
  55.09\% &
  67.77\% &
  12.32\% &
  75.01\% &
  76.95\% &
  79.44\% &
  55.77\% &
  97.90\%\\
 &
  l1 &
   56.54\% &
  43.65\% &
  66.03\% &
  15.81\% &
  52.78\% &
  {70.35\%} &
  63.64\% &
  68.35\% &
  88.33\% \\
   \midrule
 
\multirow{2}{*}{Comem} &
  l0 &
  
  \textbf{83.25
\%} &
  {\textbf{80.23\%}} &
  73.93\% &
  { \textbf{82.68\%}} &
  \textbf{78.26\%} &
  81.36\% &
  \textbf{89.72\%} &
  57.69\% &
  97.06\% \\
  
 &
  l1 &
  \underline{78.19\%} &
  \underline{ 77.24\%} &
  66.67\% &
  \underline{ 79.44\%} &
  \underline{ 75.56\%} &
  \underline{ 76.22\%} &
  63.64\%&
  69.62\% &
  91.12\% \\
    \midrule
\multirow{2}{*}{JustAsk} &
  l0 &
  77.80\% &
  69.03\% &
  {73.46\%} &
  53.62\% &
  73.91\% &
  \textbf{82.24\%} &
  84.11\% &
  62.50\% &
  \textbf{98.32\%} \\
 &
  l1 &
  62.35\% &
  54.38\% &
  64.74\% &
  43.72\% &
  51.67\% &
  64.66\% &
  62.34\% &
  67.09\% &
  85.00\% \\
   \midrule
\multirow{2}{*}{Frozen}  &
  l0 &
  81.13\% &
  75.22\% &
  74.68\% &
  63.77\% &
  77.17\% &
  81.09\% &
  88.79\% &
  60.58\% &
  97.90\% \\
 &
  l1 &
  70.61\% &
  66.55\% &
  62.18\% &
  58.60\% &
  71.67\% &
  74.33\% &
  62.34\% &
  62.03\% &
  90.56\% \\
  
    \midrule[0.1mm]
\multirow{2}{*}{Human} &
  l0 &
  - &
  84.38\% &
  {83.22\%} &
  87.12\% &
  { 88.83\%} &
  {80.13\%} &
  { 79.21\%} &
  {86.10\%} &
  - \\
&
  l1 &
  - &
  84.18\% &
  79.77\% &
  84.55\% &
  87.68\% &
  79.75\% &
  79.82\% &
  79.70\% &
  -
  
                        \\  \bottomrule
\end{tabular}%
}
\caption{A comprehensive comparison of VideoQA methods and human evaluation on BDIQA.
}
\label{level_res}
\end{table*}

\begin{table}[bth]
\centering

\scalebox{0.90}{
\begin{tabular}{cccc}
\toprule
Models    & overall \(Acc\)   & \(Acc_{bdi}\)      &  \(Acc_{p}\)   \\
\midrule
TimeSformer & {76.89\%} & {75.67\%} & {71.65\%} \\
     ResNet  & 73.94\%          & 73.89\%          & 66.75\%          \\
       ResNeXt-101   & 71.99\%          & 70.09\%          & 68.38\%\\
       
        CLIP ViT-L-14  &72.83\%&	72.90\%	&70.84\% \\
        RR & \textbf{77.84\%} &	\textbf{79.13\%}&	\textbf{71.93\%}\\
        \hline
        S3D  & 71.49\%          & 70.03\%          & 68.66\%          

       \\ \bottomrule
\end{tabular}%
}
\caption{ Results with different video backbones for Frozen.}
\label{video_representations}
\end{table}

\subsection{Improving Pre-trained Models Performance}
During analyzing visual reasoning techniques on BDIQA, we are surprised to find that the pre-trained models performed poorly. In order to transfer the prior knowledge of the pre-trained models to ours tasks, we improve the pre-trained models \footnote{Code
https://github.com/mao-yy/BDIQA.git } and propose two suggestions to solve the cognitive reasoning task. Case study for our models and baseline can be found in supplementary material.

\subsubsection{Perception is the basis of reasoning.}
As the above said, perception provides the foundation upon higher-order cognitive processes. Therefore, in visual cognitive tasks, the processing of visual inputs is indeed essential for successfully solving tasks.
 Our first idea is to replace the visual backbone.
 Yang \cite{yangzerofrozen} proposed a VideoQA pre-trained method based on freezing the visual model and bidirectional 
 language model using light trainable modules. And we replace the visual backbone of Frozen to conduct ablation experiments. We choose S3D \cite{s3dxie2018rethinking}, ResNext \cite{motion_fea_Hara2017CanS3},
ResNet\cite{restnet_He}, TimeSformer \cite{gberta_2021_ICML_timeformer} and CLIP ViT-L/14 \cite{ViT1Alexander} as the visual backbones of Frozen. 
We also follow Comem to use a simple GRN to extract appearance and motion features
with ResNext and ResNet (RR).  

As is shown in Table \ref{video_representations}, compared with S3D, video representations with appearance and motion features (RR) effectively improve the overall accuracy as well as \(Acc_{bdi}\) which also surpasses the single features with only one of appearance and motion features. Other video representations also improve on BDIQA to varying degrees.
This conclusion can also been validated on JustAsk, with a 14.24\% improvement shown in Table \ref{tab:our_results}.
However, the improvements of the visual backbone are not enough for Frozen to surpass the state-of-the-art model (Comem).

\begin{table}[tb]
\centering

\scalebox{0.90}{
\renewcommand\arraystretch{1.0}

\begin{tabular}{cccc}
\toprule
Models & overall \(Acc\)   & \(Acc_{bdi}\)      &  \(Acc_{p}\)   \\
\midrule
Frozen & 71.49\%  & 70.03\% & 68.66\% \\

+M	       &        72.22\% & 71.51\% & 68.11\%   \\
+RR & 77.81\% & \textbf{79.13}\% & 71.93\%  \\
    
+M+RR (ours)  & 78.62\% & 76.36\% & 74.36\% \\

\midrule
JustAsk  & 64.59\% & 64.29\% & 64.30\% 
 \\
     
+RR & 78.83\% & 76.76\% & 70.83\%  \\
      
+M+RR (ours)    & \textbf{81.01}\% & 78.54\% & \textbf{75.46}\%  \\
\midrule

Comem  &79.84\% & 78.73\% & 71.92\%  \\
\bottomrule
\end{tabular}%
}
\caption{
A comprehensive comparison of ours methods in Frozen and JustAsk. +RR means models with ResNet and ResNeXt-101. +M means models with memory module.
}

\label{tab:our_results}

\end{table}

\subsubsection{Reason like human.} Children’s ToM development is characterized by their ability to follow certain steps or patterns when solving problems. The process involves employing multiple steps of reasoning, starting from simpler concepts and gradually moving towards more complex ones. 
Inspired by human reasoning, some of cross-modal modules are designed to reason about complex tasks step by step such as memory network, hierarchical structure, etc. 
In order to help the performance of pre-trained models on BDIQA, it is available to incorporate mechanisms of human reasoning into pre-trained models.
Our main idea is to incorporate the memory network into pre-trained models which enables to extract videos and questions features as well as their context. 
As is shown in Table \ref{tab:our_results},  
although the memory network module doesn't contribute much improvement to the original version of Frozen, when combined with state-of-the-art visual feature techniques, it achieves a trivial enhancement of 0.81\%.
This module contributes a 2.18\% improvement for JustAsk and it exceeds the baseline by achieving an overall narrow margin of 1.17\%.

In conclusion, our approach is simple but our improvements result in the enhancement of both the JustAsk and Frozen. Specifically, JustAsk has emerged as the superior model, surpassing Comem as the second-best model.

\section{Conclusion and Future Work}
The paper introduces a new benchmark called BDIQA, which aims to explore the cognitive reasoning capabilities of VideoQA models. 
It offers tasks at different difficulty levels to assess the model's understanding of BDI, and fills a gap in existing machine ToM by including BDI joint inference and video-language data.
Using BDIQA, we evaluate several different VideoQA methods. 
After analysis, we have come to two guidelines for enhancing cognitive reasoning in VideoQA models.
Our approach is simple but does improve, and 
these guidelines likely provide recommendations or strategies for augmenting the models to improve their performance on BDIQA tasks.

Although BDIQA is not large and the complexity of the questions to be challenging for VideoQA models
, in order to provide cognitive intelligence to AI, we incorporate psychological theory into the design process, which is worth considering because it offers theoretical guidance for the development of AI learning.
The additional information provided enables the expansion of our dataset for more detailed behavioral studies with mental states.
The analysis of our experiment demonstrates that existing models cannot solve this task.
Thus, BDIQA has also presented existing researchers with insights into how to develop novel architectures specifically tailored for cognitive reasoning in VideoQA.
This is the issue we are going to face next, and we advocate 
these architectures can incorporate techniques from cognitive science, neuroscience, or other relevant fields to enhance AI’s cognition development.

\appendix
\section{Acknowledgments}
This work is supported by National Key Research and Development Program of China (2021ZD0111000/ 2021ZD0111004), the Science and Technology Commission of Shanghai Municipality Grant (No. 21511100101, 22511105901, 22DZ2229004) and Fund of the International Conference of Graduate Students of East China Normal University. 
Xin Lin is the corresponding author.

\bigskip


\end{document}